\documentstyle[aps,prl,psfig,epsfig,floats]{revtex}
\begin{document}
\draft
\twocolumn[\hsize\textwidth\columnwidth\hsize
\csname @twocolumnfalse\endcsname
\title{Theory for phonon-induced superconductivity in MgB$_2$}
\author{C. Joas$^1$, I. Eremin$^{1,2}$, D. Manske$^{1}$, and 
K.H. Bennemann$^1$}
\address{$^1$Institut f\"ur Theoretische Physik, Freie Universit\"at 
Berlin, D-14195 Berlin, Germany}
\address{$^2$Physics Department, Kazan State University, 420008 Kazan, Russia}
\date{May 18, 2001}
\maketitle
\begin{abstract}
 We analyze superonductivity in MgB$_2$
  observed below $T_c=39$ K resulting from electron-phonon
  coupling involving a mode at $\hbar \omega_1 = 24$ meV  
  and most importantly the in-plane B-B $E_{2g}$ vibration at $\hbar
  \omega_2=67$ meV.  The quasiparticles 
  originating from $\pi$- and $\sigma$-states couple strongly to
  the low-frequency mode and  the $E_{2g}$-vibrations respectively. 
  Using two-band Eliashberg theory, $\lambda_{\pi} = 1.4$ and 
  $\lambda_{\sigma} = 0.7$, we calculate the gap
  functions $\Delta^{i}(\omega,0)$ ($i=\pi$, $\sigma$). 
  Our results provide an explanation of recent tunneling experiments. 
  We get $H^{ab}_{c_2}/H^{c}_{c_2} \approx 3.9$. 
 \end{abstract}
]
\narrowtext
It is important to understand the physics of the recently observed
superconductivity in $\mbox{MgB}_2$ at $T_c \approx$39K\cite{akimutsu}. 
Recent experimental and theoretical results\cite{budko,hinks,picket,mazin} 
%
%
have shown that most likely strong electron-phonon 
interaction is the relevant mechanism for superconductivity in 
this compound. However, many questions remain. In recent Raman \cite{chen}, 
tunneling\cite{french,szabo}, photoemission\cite{takano}, and 
transport\cite{junod} studies it was noted that the experimental 
data are compatible with a multi-component superconducting gap. This is 
consistent with the large observed anisotropy of $H_{c_2}$\cite{budko1,simon}.
Inelastic neutron scattering(INS) experiments\cite{muranaka} and 
point-contact spectroscopy studies\cite{bobrov}  
reveal a peak at approximately 
24 meV in the phonon density of states in contrast
to first principles calculations\cite{stuttgart,osborn,yildirim}. 
Recently, this peak was also indirectly confirmed 
by the analysis of the temperature dependence of the 
upper critical field $H_{c_2} (T)$\cite{shulga}. 
The isostructural AlB$_2$ is not superconducting.

In this paper we analyze superconductivity in 
MgB$_2$ using Eliashberg theory assuming that it arises due to electron-phonon
coupling of the holes in the Boron planes
($\sigma$-states) to the E$_{2g}$-mode of B-B vibrations 
and due to the low-frequency mode interacting with the $\pi$-states.

In Fig. 1 we illustrate the simple physical picture for
superconductivity and the structure of MgB$_2$.  Similar to graphite,
it has a quasi 2D-like structure, but with strong covalent bonds. 
According to previous studies\cite{picket,mazin,stuttgart}, 
superconductivity occurs mainly in
the B-planes involving two-dimensional $\sigma$- and and three-dimensional 
$\pi$-states. In the first case 
the electron-phonon coupling can be estimated from
$g_{e-ph}\sim <\sigma_B | \vec{\nabla} V | \sigma_B> \cdot \, {\bf
e}$, where the polarization vector {\bf e} refers to the
E$_{2g}$-mode\cite{stuttgart} and $\sigma_B$ refers essentially 
to the in-plane wave function of Boron bonds. 
Note, we obtain g$_{e-ph} \sim \vec{\nabla} t \, 
\cdot \, {\bf e}$ ($t$ denotes the corresponding hopping matrix element) and 
expect a resulting moderate electron-phonon 
coupling like in transition metals. 

It is known from first principle calculations that
$\mbox{MgB}_2$ consists of four important phonon modes\cite{stuttgart}. 
Two modes involve vibrations of the Mg-B planes: 
the $E_{1u}$ mode (40meV)
corresponds to a sliding mode in $xy$-direction, and the $A_{2u}$ mode 
(49meV) corresponds to a vibration in $z$-direction. 
On the other hand, the other two modes involve the B-B bonds: 
the $E_{2g}$ mode (67meV) mentioned
earlier corresponds to an in-plane breathing mode, and the
$B_{1g}$ mode (87meV) refers to an out-of-plane tilting mode of Boron
atoms. 
First-principle
calculations of the electron-phonon coupling $\alpha^2F(\omega)$ in
Ref. \onlinecite{stuttgart} obtain  that the dominating part of the
pairing interaction stems from the $E_{2g}$ mode. In the non-superconducting 
AlB$_2$, this $E_{2g}$-mode lies at 
$\hbar \omega = 120$ meV. 
Recent INS and Raman
experiments in MgB$_2$ reveal an 
additional peak in the phonon density of states at approximately 
17-25meV\cite{muranaka,bobrov,lampakis} and, most importantly, its softening 
in the superconducting state. It was noted\cite{bobrov} 
that this mode very likely is involving Mg vibrations. Thus, one expects the 
coupling of this mode mainly to the $p_z$- ($\pi$) band due to its  
dispersion along the $z$-axis.  Therefore, we conclude that 
superconductivity in MgB$_2$ is mainly induced by Boron 
$\pi$- and $\sigma$-states coupled to low-frequency mode at 24meV and the 
$E_{2g}$-mode at 67meV respectively. Due to the different character of both 
states and their coupling to different phonon modes one expects the 
occurrence of superconducting gaps referring mainly to the 
$\pi$- and $\sigma$-bands. 
This is necessary as well for the analysis of 
the observed anisotropy in the upper critical field. 
%
%
%
%

For our analysis we use a two-band extension of  
Migdal-Eliashberg theory\cite{butler,kresin}.  
We approximate the
phonon spectral density $F^{i}(\omega)$ by two Lorentzians centered 
at $\hbar \omega_1 \approx$ 24 meV for the 
$\pi$-band and $\hbar \omega_2 (E_{2g})$ = 67 meV for the $\sigma$-band. 
Hence,
\begin{equation}
\alpha^2F^{i}({\bf q},\omega) = \, \frac{1}{\pi} g_{p}^{(i)}
\frac{\omega\Gamma_i^3}{\left[\left(\omega - \omega_i\right)^2
 + \Gamma_i^2\right]^2} \, F^i ({\bf q}) \, ,
\end{equation}
describing the pronounced phonon peaks at the observed frequencies.
Here, $g_{p}^{(i)}$ refers to the strength of the phonon mode 
$i$, and $\Gamma_i$
corresponds to its damping.
For simplicity we neglect the ${\bf q}$-dependence of the 
spectral functions, i.e.
$F_i ({\bf q}) \equiv 1$. We employ $\Gamma_1 = 5$ meV and 
$\Gamma_2 = 10$ meV. Thus, the Eliashberg coupling constants 
(for $\sigma$- and $\pi$-states)
\begin{equation}
\lambda_i = 2 \int_0^{\infty} d\omega \omega^{-1} \alpha^2F^{i}(\omega)
\end{equation}
become
\begin{equation}
\lambda_i =  \frac{1}{\pi} 
\frac{g_p^{(i)}}{
\left[\frac{\pi}{2} + \arctan 
\left( \frac{\omega_i}{\Gamma_i} \right) + 
\frac{\left(\omega_i / \Gamma_i \right)}
{1 + \left( \omega_i / \Gamma_i \right)^2}  \right]}
\quad .
\end{equation}
Note, the single T$_c$ is provided by the non-zero value of 
$\lambda_{\pi \sigma}$ and $\lambda_{\sigma \pi}$ due to 
transitions between $\sigma$- and $\pi$-states\cite{butler}. However,  
their effect at low temperatures on the solution of the Eliashberg 
equations seems not to be of significant importance. Thus we neglect them 
for T=0. Regarding the value of $\lambda_{\sigma}$ we have chosen it 0.7 
in quantitative agreement with first principle calculations\cite{picket}. 
The strength 
of the coupling of the $\pi$-electrons to the low-frequency mode is unknown.
However, using an Einstein model for the phonon spectrum yielding 
$\alpha^2 F(\omega) = g^2 /\omega_E$, one expects 
$\lambda_{\pi}>\lambda_{\sigma}$ \cite{verynew} (see also \cite{rainer}).
Taking into account also recent experiments indicating that the larger gap 
shows much smaller coherence peaks in the tunneling density of 
states\cite{french} one concludes also that $\lambda_{\pi}>\lambda_{\sigma}$.
In the following, we approximate $\lambda_{\pi} = 1.4$. 
Then we use Eliashberg equations referring to the $\pi$- and 
$\sigma$-states separately for T=0 ($\omega_c$ is a cutoff 
frequency):
\begin{eqnarray}
\Delta^{i}(\omega) & = &
\frac{1}{Z^{i}_s (\omega)}\int_0^{\omega_c} d\nu \,
\mbox{Re }\Bigg\{  \frac{\Delta^{i}(\nu)}{
[\nu^2 - \left(\Delta^{i}\right)^2(\nu) ]^{1/2}} \Bigg\}
\nonumber\\[1ex]
& \times &
\{K^{i}_+ (\nu, \omega) - N_{i}(\epsilon_F) U^{i}_c \}
\quad ,
\label{eq:delta}
\end{eqnarray} 
\begin{eqnarray}
\lefteqn{\left[ 1 - Z^{i}_s (\omega)\right] \, \omega = }
\nonumber\\
& &
\int_0 ^{\infty}d\nu \, \mbox{Re }\Bigg\{ 
\frac{ \nu }{ [\nu^2 - \left(\Delta^{i}\right)^2 (\nu)]^{1/2} } \Bigg\}
K^{i}_{-}(\nu, \omega)
\quad ,
\label{eq:z}
\end{eqnarray}
with the kernel
\begin{eqnarray}
K^{i}_{\pm} (\nu, \omega) & = &
\int_0 ^{\omega_{max}} 
d\omega^{'} \, \alpha^2 F^{i}(\omega')
\nonumber\\
& \times &
\Bigg( \frac{1}{\omega +\omega^{'} +\nu +i\delta } \mp
\frac{1}{\omega -\omega^{'} -\nu +i\delta }
\Bigg)
\quad,
\label{eq:k}
\end{eqnarray}
where $i$ refers to $\pi$- and $\sigma-$electrons.
Here $Z^{i}_s(\omega)$ denotes the effective mass renormalization function 
of the $\sigma$- and $\pi$-bands in the superconducting state 
and $U^{i}_c$ refers to the screened Coulomb (pseudo-)potential calculated 
within the static Thomas-Fermi approximation for 3D metals. 
Solving these equations 
self-consistently for the given phonon spectrum ($\pi$-states couple 
to $\omega_1$, while $\sigma$-states to $\omega_2$-mode) we obtain 
$\Delta^{i}(\omega,T)$ with structures due to the two pronounced modes in 
$\alpha^2 F^{i}(\omega)$\cite{led}. Regarding the choice of the values for 
$\mu^{*}_i$,
note that the $\sigma-$band is more metallic than the 
$\pi$-band, which is reflected by their different 
Fermi velocity and density of states at the Fermi level \cite{picket,mazin}.
This leads to $\mu^{*}_{\sigma} > \mu^{*}_{\pi}$. 
Taking $\mu^{*}_{\pi} = 0.1$ and the appropriate values of 
$N_i(\epsilon_F)$ and $\omega_i$, we obtain
$\mu^{*}_{\sigma} = 0.18$. These values of $\mu^{*}_{i}$ and 
$\lambda_i$ give the observed magnitude of the superconducting gaps, 
their ratio and the anisotropy of the upper critical field $H_{c2}$.
From this point 
of view, MgB$_2$ can therefore be referred to as a conventional 
electron-phonon mediated superconductor with moderate coupling.

In Fig.\ref{fig2} we show the results for the superconducting gaps for the 
$\sigma$- and $\pi$-bands 
solving Eqs. (\ref{eq:delta})-(\ref{eq:k}). 
Due to the larger $\lambda$ and smaller $\mu^{*}$ 
the gap value $\Delta^{\pi}_0 = 7.5$ meV 
for the $\pi$-band is almost two times larger than superconducting gap 
$\Delta^{\sigma}_0 = 4.2$ meV for the $\sigma$-band. The 
values of the gaps and their ratio are in good agreement with experimental 
data\cite{chen,french,szabo,junod}. 
The curves also reflect the underlying 
phonon spectrum $\alpha^2 F^{i}(\omega)$. Re $\Delta^{i}(\omega)$ 
changes its sign above $\omega_i$ due to the over-screened pairing 
potential and Im $\Delta^{i}(\omega)$ starts to increase at 
$\omega_i$ because of phonon emission processes.

In order to test whether the effective one-band model can be applied, 
we also have solved the one-band version of the Eliashberg equation 
for the average $\bar{\lambda}=\frac{N_{\sigma}(0)}{N_{tot}}\lambda_{\sigma}+
\frac{N_{\pi}(0)}{N_{tot}}\lambda_{\pi} \approx 0.87$ and average 
$\bar{\mu}^{*} = 0.5\left(\mu_{\sigma}^{*}+\mu_{\pi}^{*}\right)$. The 
corresponding phonon density of states includes both low-frequency and 
$E_{2g}$ peaks with the coupling to the effective Bloch states. The values 
of $g_i$ were chosen in order to have the same relative strength  
as in the case of the two-band model and also to satisfy the absolute 
value of $\tilde{\lambda}$. Then, we find a superconducting gap 
$\tilde{\Delta}_0 = 4$ meV.
%
%
%
%
%
%
%

In Fig. \ref{fig3}(a) 
we present our results for the tunneling density of states for the 
effective gap. The obtained coherence peaks at 4 meV due 
to the effective superconducting gap $\tilde{\Delta}_0$ 
agree with the experimentally observed 
position of the order of 4 meV\cite{french}. 
However, no further pronounced 
structure at approximately 7 meV as observed in experiment is obtained. 
This suggests that an effective one gap approximation is not a good one.

In Fig.\ref{fig3}(b), we show results for the two-band Eliashberg theory. 
The tunneling density of states resulting from $\pi$- and $\sigma$-bands
is then given by
\begin{eqnarray}
N_S(\omega)/N_n(\epsilon_F) & = & 
\frac{N_{\pi}(0)}{N_{tot}}\mbox{Re}\left[\frac{\omega}
{\sqrt{\omega^2 - (\Delta^{\pi} (\omega))^2}}\right] + \nonumber\\ 
&+& \frac{N_{\sigma}(0)}{N_{tot}}\mbox{Re}\left[\frac{\omega}
{\sqrt{\omega^2 - (\Delta^{\sigma} (\omega))^2}}\right].
\label{tunnel}
\end{eqnarray}
The results depicted in Fig.\ref{fig3}(b) clearly show larger and 
smaller coherence peaks reflecting the corresponding 
superconducting gaps in the $\pi$- and $\sigma$-bands respectively. 
The position 
of these peaks($\Delta^{\sigma}_0 = 4.2$ meV and $\Delta_0^{\pi} = 7.5$ meV) 
are 
in good agreement with experimental results (4.2 meV and 7.5 meV) indicating 
that these peaks may result from two superconducting gaps. 
The smaller coherence peak of the 
larger gap is due to the smaller density of states of the $\pi$-band at the 
Fermi level with respect to the $\sigma$-states (see Eq. (\ref{tunnel})). 
This is another indication 
that the larger gap has to be attributed to the $\pi$-states rather than to 
a $\sigma$-band. 

In order to support further the validity of our results, 
let us now turn to the discussion of the resulting anisotropy of 
the upper critical field H$_{c2}$. 
According to the anisotropic Ginzburg-Landau theory equations, the 
upper critical field along the $c$-axis is given by 
$H_{c_2}^{c}=\frac{\Phi_0}{\pi\xi^{2}_{ab}}$ and for the magnetic field 
applied in the $ab$-plane by 
$H_{c_2}^{ab}=\frac{\Phi_0}{\pi\xi_{ab}\xi_{c}}$. 
Here, $\Phi_0$ is the superconducting flux quantum, 
$\xi_{ab}$ and $\xi_{c}$ are the coherence lengths referring to 
the $ab$-plane and the $c$-axis in the clean limit, respectively. 
The coherence length 
is determined as $\xi \approx \frac{\hbar v_F}{\pi \Delta_0}$. Thus, 
one gets the anisotropy ratio 
\begin{equation}
\frac{H^{ab}_{c_2}}{H^{c}_{c_2}} = \frac{v^{ab}_F 
\Delta^{c}_0}{v^{c}_F \Delta^{ab}_0}
\quad.
\end{equation} 
Since only the $\pi$-band 
has a substantial dispersion along the $k_z$-axis, we can approximate 
$v^{c}_F$ by $v^{z}_F \approx 4 \times 10^7$cm/s for 
the $\pi$-band\cite{mazin}. Similar arguments for the 
$\sigma-$band lead to 
$v^{ab}_F \approx v^{xy}_F \approx 8.5 \times 10^7$cm/s \cite{picket}. 
Substituting our values for the superconducting 
gaps we obtain $\frac{H^{ab}_{c_2}}{H^{c}_{c_2}} \approx 3.9$, which is in 
good agreement with the experimental observation\cite{budko1,simon}. 
The recent suggestion \cite{Brinkman,Golubov} that the larger gap belongs to 
the $\sigma-$band and the smaller gap occurs in the 
$\pi$-band reduces significantly the anisotropy of the upper 
critical field and also leads to the intensity ratios of the 
coherence peaks in the tunneling density of states being 
inconsistent with experiment \cite{french}. However, we get 
$H^{ab}_{c_2} \approx 5.6 Tesla$ taking into account 
the renormalization of the Fermi velocities due to 
electron-phonon interaction, while some experiments give 15$Tesla$ 
or even large values. Note, further correcting the LDA results for $v_F$ by a 
factor 1.5 for the $\pi$- and $\sigma-$states we get 
$H^{ab}_{c_2} \approx 14.6 Tesla$. Many-body effects like 
electron-electron correlations may cause this correction. 
%
%
%
%
%
%
%

In Fig.\ref{fig4} we show the calculated effective 
mass renormalization functions 
$Z^{i}_s(\omega)$ for the $\sigma$- and $\pi$-quasiparticles reflecting 
many-body effects due to electron-phonon interaction and 
which describe the strength of the interaction between 
the corresponding quasiparticles. Their structures again reflect the 
underlying phonon spectrum $F^{i}(\omega)$ at 24 meV for the $\pi$-states 
and at 67 meV for the $\sigma$-electrons. 
Remarkably, the mass renormalization 
for $\pi$-electrons is larger due the difference in the corresponding 
$\lambda_i$.

Let us now estimate the high superconducting transition 
temperature and the corresponding isotope effect. As we already mentioned 
the single T$_c$ is expected due to non-zero values of 
$\lambda_{\pi\sigma}$. Moreover, as was shown earlier \cite{butler}, 
even in a two-band case one still can use effective parameters for the 
determination of T$_c$.
Thus, we use the McMillan 
formula ($\omega_2 \sim 1/\sqrt{M_B}$)\cite{mac}
\begin{equation}
k_B T_c \approx \frac{\hbar \omega_{D} }{1.2} \exp \Bigg\{
- \frac{ 1.04 (1+\bar{\lambda}) }{\bar{\lambda} - \bar{\mu}^{*} 
(1+0.62\bar{\lambda}) }
\Bigg\} 
\quad , 
\label{eq5}
\end{equation}
with $\omega_D \approx \omega_2$, and an 
effective $\bar{\lambda} = 0.87$ and $\bar{\mu}^{*} = 0.14$ resulting 
in T$_c \approx 32$K. We get for isotope coefficient 
$\alpha = -d\ln T_c / d \ln M_B$ the 
value 0.43. Assuming that the isotope effect will be mainly determined 
by the coupling between $\sigma$-states and the E$_{2g}$-mode we should 
use in Eq. (9) for $\lambda$ and $\mu^{*}$ the values referring to the 
$\sigma$-states. Thus, we find the absolute value $\alpha \approx$0.23 
which is in agreement with experiment\cite{budko}.  
We also would like to point out that despite 
of the larger gap in the $\pi$-band, superconductivity
($n_s$)
and T$_c$ are mainly 
due to the $\sigma-$states. 
This can be seen from the estimation 
of the superfluid density 
$n^{i}_s \sim N_i(\epsilon_F) \omega_{i}/$(unit volume) 
which is much larger for the $\sigma$-electrons due to their larger density of 
states at the Fermi level and the large $\omega_2$.

In summary, we can explain important facts about superconductivity in
MgB$_2$ as resulting from electron-phonon coupling of the 
$\pi$-states and the $\sigma$-states with the corresponding phonon modes. 
The larger gap results from the $\pi$-band coupled strongly to the 
low-frequency mode, while the 
smaller gap is due to the 
coupling of $\sigma$-bands with the E$_{2g}$-mode. 
The structure seen by recent tunneling 
experiments is due to the larger superconducting 
gap of the $\pi-$band. We also find $H^{ab}_{c_2}/H^{c}_{c_2} \approx 3.9$ in 
good agreement with experiment. 
We safely 
conclude that there is no need for invoking 
other pairing mechanisms besides electron-phonon interaction.
Direct calculations of $g^{i}_{el-ph}$ and $\lambda_i$  are
necessary for a further support of our simple model. In particular,  
the origin of the 24 meV peak observed by INS and Raman scattering 
must be clarified.  
Using our picture for superconductivity in MgB$_2$, one may  
estimate the absence  of superconductivity in AlB$_2$ 
due to the decrease of $\lambda_{\sigma}$ ($\lambda \sim |g|^2 /\omega_{2}$) 
and the shift of $\sigma$-states below $\epsilon_F$\cite{picket}. 

We thank K. Scharnberg, D. Fay, R. Kremer, K.D. Schotte for 
helpful discussions and, in particular, S.-L. Drechsler for continuous 
interest in our work. 
The work of I. E. is supported by the 'Alexander von Humboldt Foundation'.

\begin{figure}[t]
\centerline{\psfig{clip=,file=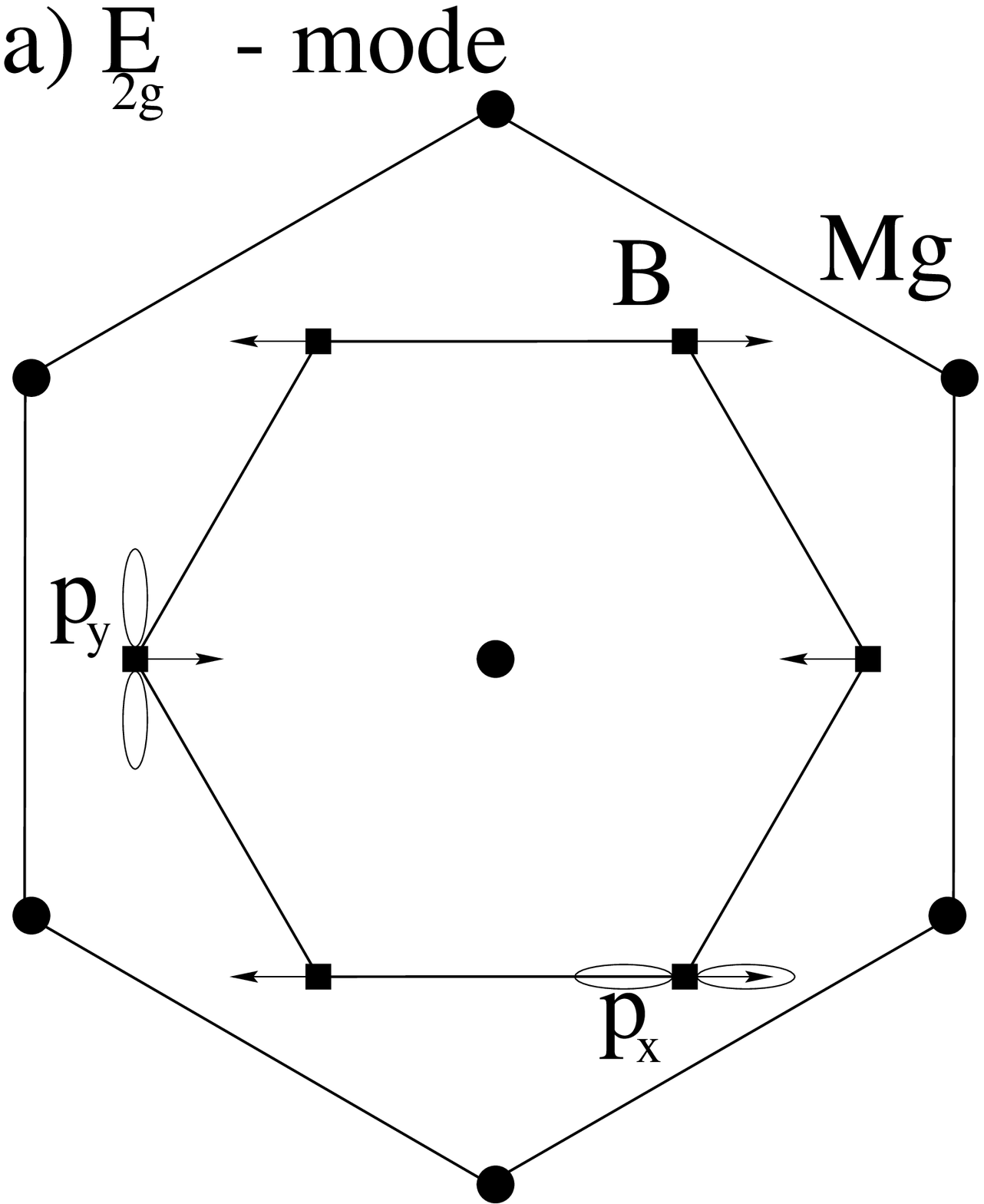,width=4.2cm,angle=0}}
\vspace*{0.5cm}
\centerline{\psfig{clip=,file=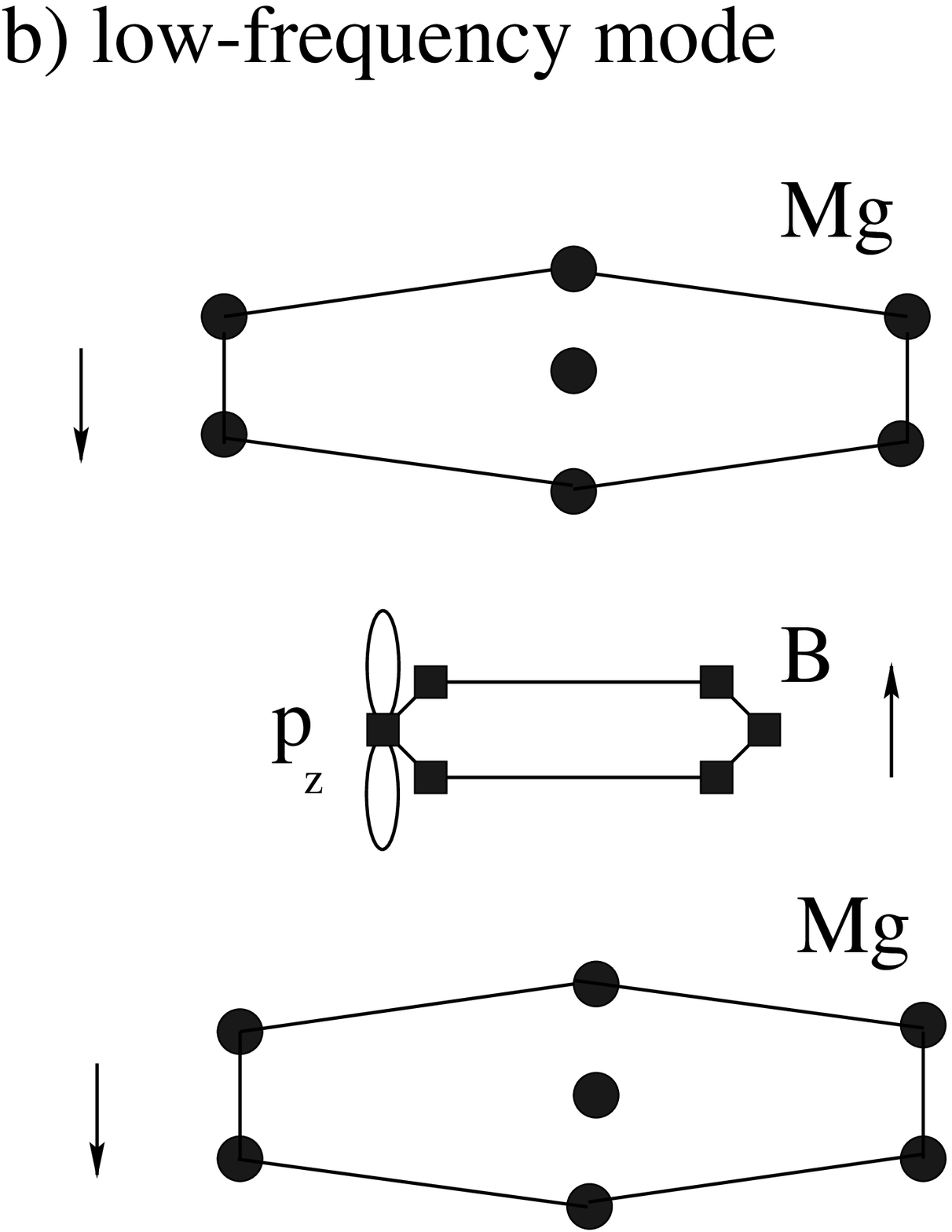,width=4.2cm,angle=0}}
\caption{Simplified physical picture for superconductivity in MgB$_2$:
a) Top view of the elementary unit cell of MgB$_2$ 
consisting of hexagonal layers of Mg and graphite-like Boron layer. The 
Boron $\sigma$-states ($p_x$ and $p_y$) are involved in 
superconductivity and couple to the $E_{2g}$ mode of B-B vibrations shown 
by arrows; b) the Boron $\pi$-states ($p_z$) are 
coupled to the out-of-plane vibrations shown schematically by arrows.}
\label{fig1}
\end{figure}
\begin{figure}[t]
\hspace*{0.2cm}
%
\vspace*{2ex}
\caption{ Results for the gap function $\Delta^{i}(\omega)$ 
at $T=0$K using two-band Eliashberg-theory: (a) $\sigma$-band with  
$\lambda_{\sigma} = 0.7$ and  
$\mu_{\sigma}^{*}=0.18$ yielding $\Delta^{\sigma}_0 = 
\Delta^{\sigma}(\omega = 0) = 4.2$ meV. 
(b) $\pi$-band with $\lambda_{\pi} = 1.4$ and  
$\mu_{\pi}^{*}=0.1$ yielding $\Delta^{\pi}_0 = 7.5$ meV.}
\label{fig2}
\end{figure}
\begin{figure}[t]
%
%
\hspace*{0.2cm}
%
%
\vspace{0.5ex}
\caption{Calculated tunneling density of states: (a) using an 
effective one-band Eliashberg theory with elastic 
scattering strength $\gamma = 0.3$meV; 
(b) using two-band Eliashberg theory and the same $\gamma$ as in (a). 
In the inset, experimental data is displayed \protect\cite{french}.}
\label{fig3}
\end{figure}
\begin{figure}[t]
%
%
%
%
\vspace{2ex}
\caption{Results for the mass renormalization function $Z^{i}_s(\omega)$ 
at $T=0$K using two-band Eliashberg-theory for (a) the $\sigma$-band and 
(b) the $\pi$-band.}
\label{fig4}
\end{figure}


\begin{references}
\bibitem{verynew}We estimate $\lambda_{\pi}/\lambda_{\sigma} \propto 
\frac{\omega_2}{\omega_1}\frac{N_{\pi}(0)}{N_{\sigma}(0)}
\vert \frac{<\pi | \vec{\nabla} V | \pi >}
{<\sigma | \vec{\nabla} V | \sigma>}\vert^2$. 
Assuming $<\pi| \vec{\nabla} V |\pi> \,\, 
\ge 1.4<\sigma|\vec{\nabla} V|\sigma>$ due 
to more localized character of the $\pi$-states, one might get 
$\lambda_{\pi} = 2 \lambda_{\sigma}$.
\end{references}
\end{document}